# High-resolution time-to-digital converters (TDCs) with a bidirectional encoder

Yu Wang    Wujun Xie    Haochang Chen    David Day-Uei Li

**Abstract:** A high-resolution time-to-digital converter (TDC) based on wave union (four-edge WU A), dual-sampling, and sub-TDL methods is proposed and implemented in a 16-nm Xilinx UltraScale+ field-programmable gate array (FPGA). We combine WU and dual-sampling techniques to achieve a high resolution. Besides, we use the sub-TDL method and the proposed bidirectional encoder to suppress bubbles and encode four-transition pseudo thermometer codes efficiently. Experimental results indicate the proposed TDC achieves a 0.4 ps resolution with a 450MHz sampling clock and a 3.06 ps RMS precision in the best-case scenario. These characteristics make this design suitable for particle physics, biomedical imaging (such as positron emission tomography, PET), and general-purpose scientific instruments.

## I. Introduction

Time-to-digital converters (TDCs) are high-precision time sensors converting a time interval (TI) into a digital code. They are widely used in time-resolved applications, such as particle physics [1–3], positron emission tomography (PET) [4,5], random number generation [6,7], Raman spectroscopy [8,9], and light detection and ranging (LiDAR) [10,11].

We can use application-specific integrated circuits (ASICs) or field-programmable gate arrays (FPGAs) [12] to implement TDCs. FPGA-TDCs are widely applied in scientific research and prototype designs, as they have a shorter development cycle and lower development costs than ASIC-TDCs. Most FPGA-TDCs use cascade carry chains to form the tapped delay line (TDL) since carry chains are standard in modern

FPGAs and have dedicated routing resources.

The resolution (the minimal detectable TI, usually called the least significant bit, LSB) is the primary metric for a TDC. For TDL-TDCs, it is determined by the carry chain's propagation delay and can be defined as:

$$Q = \frac{T}{n}, \tag{1}$$

where *T* is the period of the sampling clock and *n* is the number of time bins. Although TDL-TDC's resolution is improved with the advances in complementary metal-oxide-semiconductor (CMOS) manufacturing technologies, the achievable resolution is still limited. Architectures, such as Vernier [13] and multi-chain merging [14], have been proposed to break this process-related limitation. However, the Vernier architecture suffers from deadtime, and the multi-chain architecture consumes significant hardware resources. Hence, these two architectures have a trade-off between the resolution and other metrics.

In 2008, Wu and Shi proposed wave union (WU) TDCs, respectively achieving 25 ps (WU A) and 10 ps (WU B) precisions in a Cyclone II FPGA [15]. The concepts of these two methods are similar to the multi-chain merging architecture, utilizing multi measurements for the same TI to achieve a better resolution and precision. However, in these two WU methods, multi-measurements are conducted by feeding a train of pulses (including rising and falling edges) rather than implementing parallel TDLs. Hence, WU methods are more effective in hardware utilization than multi-chain merging. The difference between WU A and WU B lies in the length of the pulse train. With a no-feedback wave launcher, WU A generates a finite-length pulse containing a limited number of logic transitions. However, WU B generates an infinite-length pulse train with a feedback wave launcher. Therefore, WU B has a better resolution than WU A but with a longer dead time [15]. Since the WU method was proposed, it has been the mainstream for high-resolution TDL-TDCs. For example, Wang *et al.* implemented a 4-snapshot TDC (WU B) in a Virtex-4 FPGA [16], Szplet *et al.* implemented a 6-edge

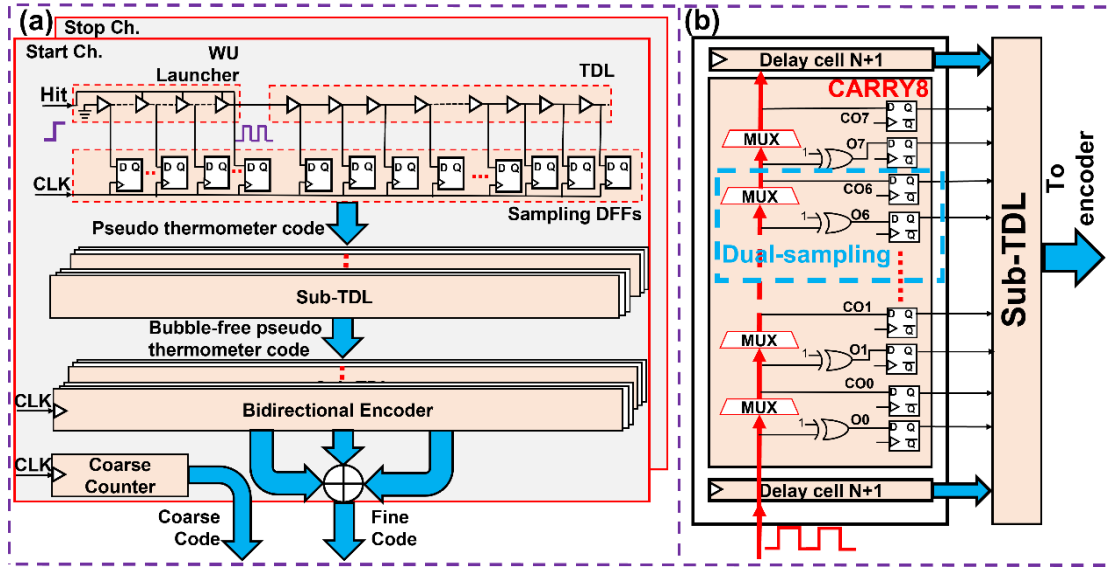

Fig. 1. The block diagram of the (a) proposed TDC system, and (b) dual-sampling method with CARRY8.

TDC (WU A) in a Spartan-6 FPGA [15], and Xie *et al.* implemented a 2-edge TDC (WU A) in an UltraScale FPGA [18].

However, to our knowledge, no WU TDC is implemented in 16-nm FPGA devices. Hence, in this work, we designed and implemented a 4-edge WU TDC (WU A) in a 16-nm UltraScale+ MPSoC device. We combined the dual-sampling and WU methods to achieve a high resolution. Besides, we also used the sub-TDL method to suppress bubbles and the proposed bidirectional encoder to encode bubble-free four-transition pseudo thermometer codes.

The remainder article is structured as follows. Section II describes the architecture and design of the proposed TDC, Section III presents the experimental results, and Section IV compares the proposed TDC with other WU TDCs. Finally, Section V concludes this work.

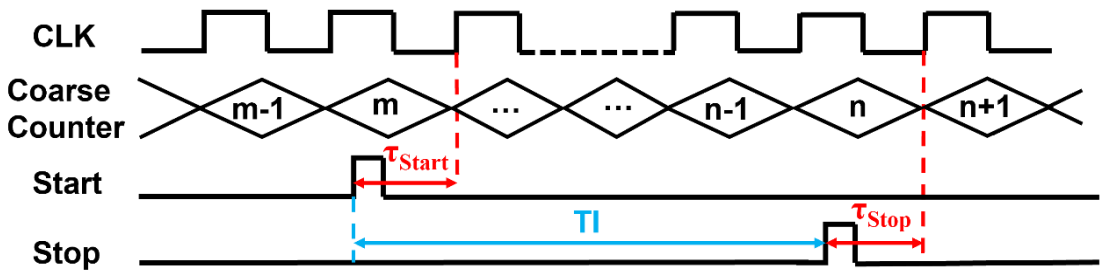

Fig.2. The timing diagram of measurement.

## II. Architecture and Design

The TDC system is shown in Fig. 1a. It consists of start and stop channels (Start Ch. and Stop Ch. in Fig 1a). These two channels are driven by the same clock and responsible for recording starting and stopping timestamps for a TI. According to the timing diagram shown in Fig. 2, the starting and stopping timestamps can be respectively defined as:

$$timestamp_{start} = \text{m} \times T - \tau_{start}, \qquad (2)$$

and

$$timestamp_{stop} = \text{n} \times T - \tau_{stop}, \qquad (3)$$

where $T$ is the period of the coarse counting clock, $m$ and $n$ are coarse codes output from the coarse counter, and $\tau_{start}$ and $\tau_{stop}$ are time intervals corresponding to respective fine codes. Hence, the TI can be calculated as:

$$TI = timestamp_{stop} - timestamp_{start} = (\tau_{start} - \tau_{stop}) + (n - m) \times T. \qquad (4)$$

A binary or gray code counter can easily achieve the coarse counter. Here, we focus on fine-time interval measurements in this paper.

### A. CARRY8 and Dual-sampling

When a hit comes, the WU launcher generates the pulse train in each channel and feeds it into the TDL. Then, when a rising edge of the sampling clock comes, the TDL's outputs are registered by D flip-flops (DFFs) to evaluate the time interval between rising edges of the hit and the sampling clock ($\tau_{Start}$ and $\tau_{Stop}$ in Fig. 2).

Hence, as the delay cell, the CARRY8 is the backbone for TI measurements. Implemented by CARRY8s, the dual-sampling method was firstly proposed by Wang and Liu [19]. As shown in Fig. 1b, the TDL (highlighted in red) is based on cascaded

CARRY8s, and each CARRY8 contains eight multiplexer (MUX)-based delay elements [20]. Like its predecessors (CARRY4s), each delay element in CARRY8 has two outputs (O and CO in Fig. 1b). Differently, in each slice, these two outputs can be simultaneously sampled by DFFs in the CARRY8 rather than either "O" or "CO" sampled in the CARRY4 [21]. Therefore, with this method, eight delay elements can output sixteen taps, equivalent to subdividing one bin into two bins. Thus, the resolution is improved without increasing the system complexity.

## B. Sub-TDL and Wave Union Launcher

Although the TDL's outputs are sampled by DFFs, they cannot be encoded directly as there are unexpected logic transitions (for example, unexpected "0" among "1"s or "1" among "0"s ), usually called bubbles. Bubbles are caused by clock skews and TDLs' uneven propagation delays [22], and they cause coding errors. The Wallace tree encoder [23], bin realignment [24], and ones-counter [25] were proposed to resolve bubble problems. However, the Wallace tree encoder and ones-counter increase hardware utilization, and iterations for bin realignment are time-consuming. Hence, these methods are unsuitable for a TDC with many time bins.

To resolve bubble problems without extra design complexity, we used the sub-TDL method [22] (or the decomposition method [26]) in our design. The sub-TDL architecture is shown in Fig. 3. The bin-width of the dual-sampling TDL is highlighted in yellow, and the sub-TDL's bin-width is highlighted in blue. The sub-TDL elongates time intervals between taps by decomposing TDL's sampling taps to minimize the impact of clock skews and obtain bubble-free outputs. Then all subsets (results from sub-TDLs) are summed and interpolated to maintain the TDC's resolution. The number of sub-TDLs depends on the maximal bubble depth (MBD). In Ref. [22], there are four sub-TDLs for a Virtex-7 FPGA and eight sub-TDLs for a Kintex UltraScale FPGA. However, in the UltraScale+ MPSoC device, the observed MBD is sixty for the dual-sampling TDL. Hence, we build sixty-four Sub-TDLs for our design.

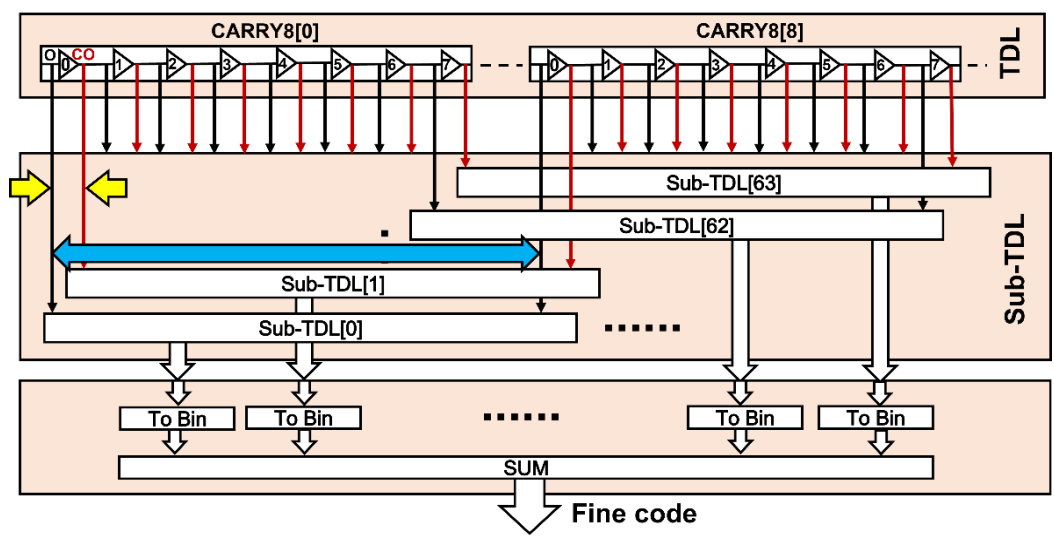

Fig. 3. The architecture of Sub-TDL.

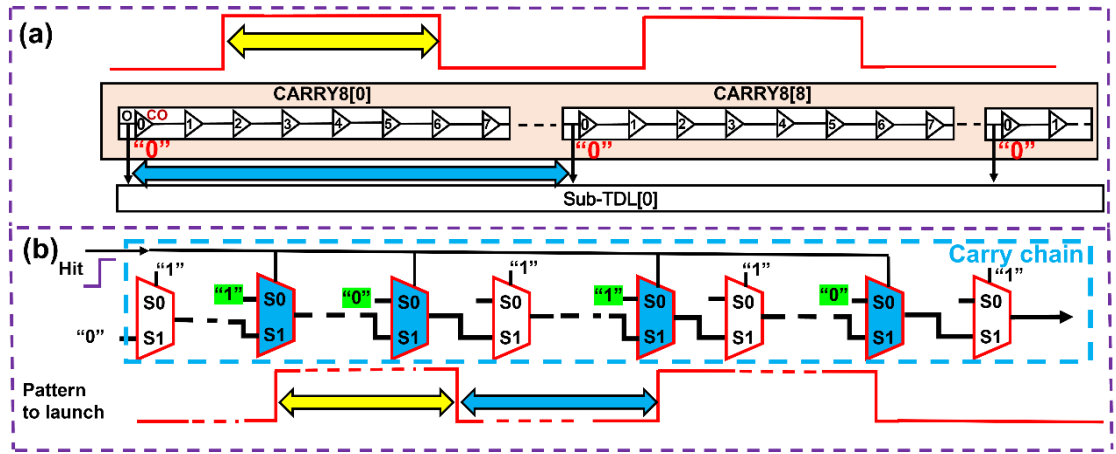

Fig. 4. (a) The concept of undetectable wave patterns in the Sub-TDL. (b) The block diagram of wave union (WU A) launcher.

However, as shown in Fig. 4a, when the WU method is combined with sub-TDLs, logic transitions may be undetectable in sub-TDLs if the pulse width (highlighted in yellow in Fig. 4a) is shorter than sub-TDL's bin-width (highlighted in blue in Fig. 4a). To avoid this and precisely control the pulse width, we implemented the WU launcher with CARRY8s (shown in Fig. 4b). By using the input signal ("hit" in Fig. 4b) as the MUX-based delay elements' select signal, the WU launcher can work in the "Standby Mode" and "Launch Mode". When the input is "0" (low-logic level), the WU launcher works in the "Standby Mode" and stores the wave pattern in the TDL. Then the stored pattern is launched when the input changes to "1" (high-logic level). The stored pattern is configured by the "S0" input (highlighted in green in Fig. 4b) of the delay element. The pulse width is configured by the number of delay elements among adjacent "configuring elements" (highlighted in blue in Fig. 4b). Our design configures the wave pattern as "01010" to contain four logic transitions. In the designed WU launcher, the width of the positive pulse (highlighted in yellow in Fig. 4b) is configured as eighty taps in the dual-sampling TDL. However, rising edges propagate faster than falling edges in the TDL, causing the width of the negative pulse (highlighted in blue in Fig. 4b) to decrease when a pulse train propagates along the TDL [18]. Hence, we configure the width of the negative pulse as one hundred and twelve taps in the dual-sampling TDL to ensure the negative pulse is always detectable in every sub-TDL.

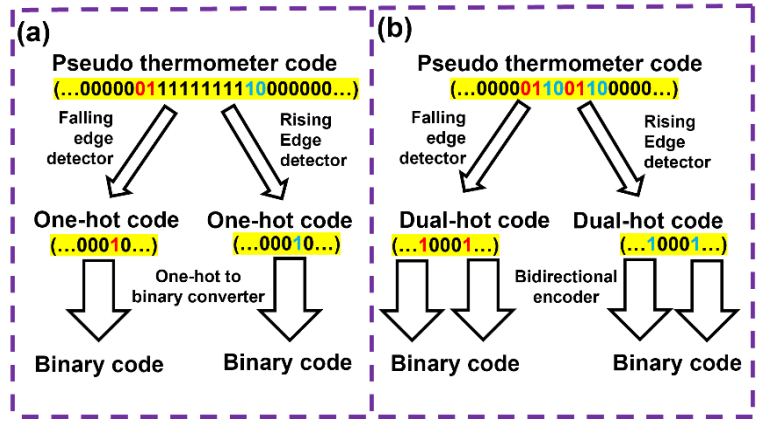

Fig.5. Encoding workflow of (a) two-edge WU TDC and (b) four-edge WU TDC.

## C. Bidirectional Encoder

With the well-designed WU launcher, four logic transitions (two rising edges and two falling edges) are detectable in every sub-TDL. In our previous work (a TDC with two-edge WU A) [27], rising and falling edges in sub-TDLs' outputs are respectively detected and converted to one-hot code by positioning "0-1" and "1-0" patterns. Then every one-hot code is converted to the corresponding binary code for final result calculations. However, that strategy is out-of-work in the proposed TDC since there are more than one rising and falling edges in every sub-TDL's output (the comparison is shown in Fig. 5). Hence, we propose a bidirectional encoder for the four-edge WU A TDC.

The bidirectional encoder is shown in Fig. 6a. It consists of rising-edge and falling-edge one-hot code generators, responsible for converting a four-transition pseudo thermometer code to four one-hot codes. Both one-hot code generators contain a pattern detector and an edge detector. When a pseudo thermometer code comes, the pattern detectors and edge detectors locate logic transitions, respectively, and then one-hot codes are generated according to logic transition positions. For example, in the rising-edge one-hot code generator, the pattern detector locates the pattern "100000", whereas the edge detector detects the rising edge "10". Then the pattern detector can generate a one-hot code since the width of the negative pulse (highlighted in green in Fig. 7a) is precisely controlled to less than 5 taps in every sub-TDL. However, two "10" patterns can be found in the pseudo thermometer code. Hence, the pattern detector's output is connected with the edge detector, and an XOR is conducted between the one-hot code from the pattern detector and the dual-hot code from detecting the pattern "10" (shown in Fig. 7a). Then, the other one-hot code for rising edges is generated by the edge detector. The falling-edge one-hot code generator is similar to the rising-edge one-hot code generator. Differently, for the falling-edge detection, the pattern detector identifies the pattern "000001", and the edge detector identifies the pattern "01". Then, similar to our previous work [27], we can convert four one-hot codes to four binary codes to generate the final code. Besides, Fig. 7b shows the encoding flow for rising edges when the width of the negative pulse (highlighted in green in Fig. 7b) is more than 5 bits in a sub-TDL. The width exceeding the limit (5 taps in sub-TDLs) causes the pattern detector to fail, leading to incorrect one-hot codes from the pattern and edge detectors.

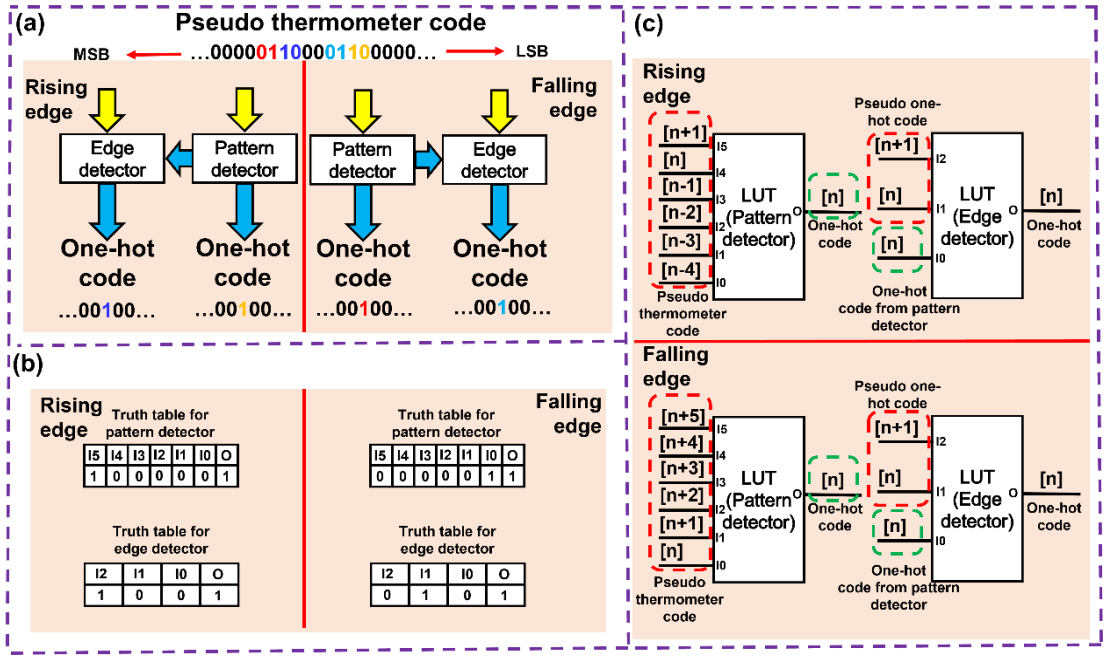

Fig.6. (a) Block diagram of the bidirectional encoder. (b) Truth tables for the bidirectional encoder. (c) Hardware implementation of the bidirectional encoder.

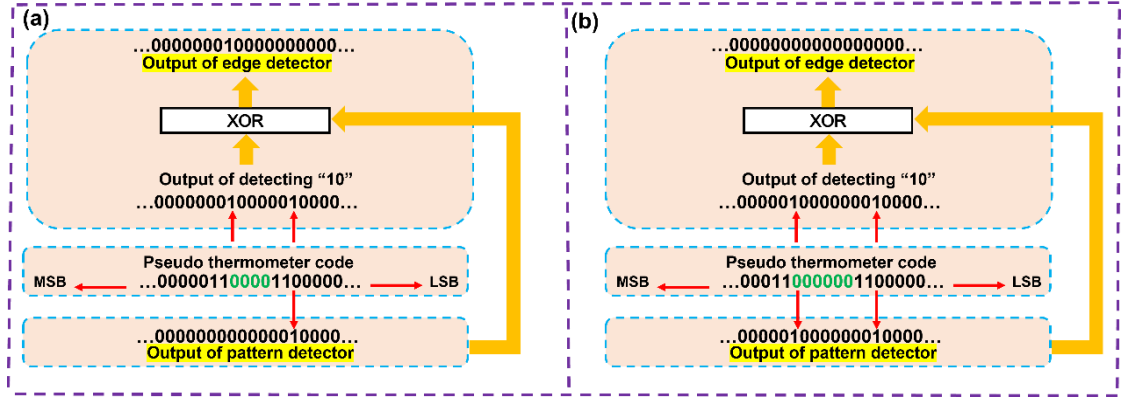

Fig. 7. Encoding flow for rising edges when the negative pulse is (a) less than 5 bits and (b) more 5 bits in a sub-TDL.

The hardware implementation of the bidirectional encoder is shown in Fig. 6c. We used LUTs to implement edge and pattern detectors but configurated them differently for rising and falling-edge generators. In our design, all LUTs for the bidirectional encoder are instantiated by Vivado Primitive [28], and the truth tables are shown in Fig. 6b.

## III. Experimental Results

We implemented the proposed TDC in the ZCU104 evaluation board [29] and closely placed two channels (Start Ch. and Stop Ch. in Fig. 1a) to reduce the offset. To evaluate TDC's performances, we used an SRS CG-635 as an external signal source. The same signal was simultaneously fed into two channels to reduce measurement errors and jitters from cables connecting the signal source and the evaluation board. For code density tests [30], this input can be treated as a random hit for two channels since it is asynchronous with the TDC's sampling clock. We used the input signal's period as a TI for RMS precision tests. Meanwhile, we can also measure the offset between two channels by calculating the difference between the same edge's timestamps recorded by two channels. The sampling clock (also known as the system clock) was sourced from an onboard crystal oscillator (IDT-8T49 [29]), and the frequency of the sampling clock was configured as 450MHz.

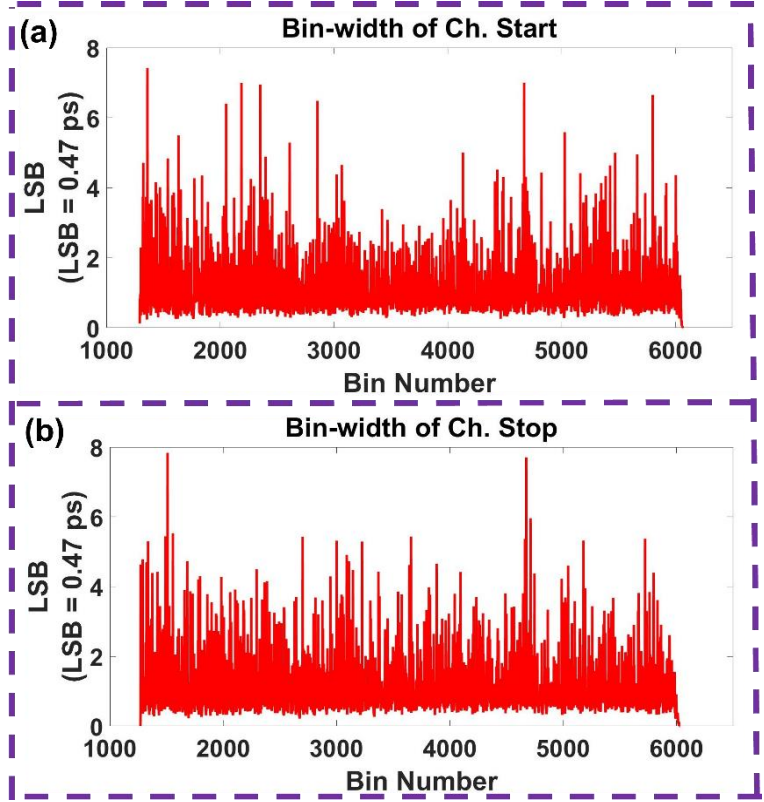

Fig.8. Bin-width of (a) the start channel and (b) the stop channel.

## A. Bin-Width and Linearity

The bin-width is the quantization step of each time bin. One million random hits (with 99.7777777MHz frequency) were fed into two channels for code density tests to estimate the TDC's bin-width. According to the number of hits collected at the $k$th bin ($n_k$), the bin-width can be estimated from:

$$W[k] = T \times \frac{n_k}{N} \qquad (5)$$

where $N$ is the number of random hits. However, suffering from clock skews and mismatches, the widths of time bins differ. These differences lead to a transfer curve rather than the desired linear quantization steps [12]. A TDC's linearity can be characterized by differential nonlinearity (DNL) and integrated nonlinearity (INL) as:

$$DNL[k] = W[k] - Q, \quad (6)$$

and

$$INL[k] = \sum_{j=0}^{k} DNL[j]. \quad (7)$$

Besides, Wu also proposed the equivalent bin-width ($\omega_{eq}$) and its deviation ($\sigma_{eq}$) to evaluate the TDC's linearity [31]. They are defined as:

$$\sigma_{eq}^2 = \sum_{i=1}^{n} \left( \frac{W[i]^2}{12} \times \frac{W[i]}{W_{total}} \right) \quad (8)$$

and

$$\omega_{eq} = \sigma_{eq} \times \sqrt{12} = \sqrt{\sum_{i=1}^{n} \frac{W[i]^3}{W_{total}}}, \quad (9)$$

where $W_{total} = \sum_{i=1}^{n} W[i]$.

The proposed TDC's bin-width is shown in Fig. 8. In both channels, the first valid time bin (not a zero-width bin) appears at around the $1250^{th}$ bin, caused by the WU launcher. A part of the carry chain constructs the WU launcher, and the pattern is generated and stored before a hit comes. Hence, the TDC's output is not zero, although there is no input and the output increases when an input hit appears. Besides, a cluster of narrow bins appears at the tail of valid time bins in both channels, caused by clock jitters. All of the above parameters for both channels are summarized in TABLE I.

TABLE I
PERFORMANCES OF THE PROPOSED TDC

| Ch. Start | | Ch. Stop | |
|---|---|---|---|
| LSB (fs) | 465 | LSB (fs) | 466 |
| DNL (LSB) | [-0.99,6.42] | DNL (LSB) | [-1,6.84] |
| INL (LSB) | [-8.79,51.56] | INL (LSB) | [-2.57,72.55] |
| $\omega_{eq}$ (ps) | 1.81 | $\omega_{eq}$ (ps) | 1.85 |
| $\sigma_{eq}$ (ps) | 0.52 | $\sigma_{eq}$ (fs) | 0.53 |

## B. Time Interval Tests

The RMS precision represents the measurement uncertainty introduced by jitters and quantization errors [30]. It is evaluated by the standard deviation ($\sigma$) of repetitive measurements for a fixed TI and improved by the bin-by-bin calibration [32]. They are respectively calculated as:

$$\sigma^2 = \sum_{i=1}^{N_T} \frac{(x_i - \mu)^2}{N_T - 1}, \tag{10}$$

and

$$t_k = \frac{W[k]}{2} + \sum_{j=0}^{k-1} W[j], \tag{11}$$

where $x_i$ is the $i$th measurement, $\mu$ is the average value for $N_T$ measurements when the TI is constant, and $t_k$ is the calibrated timestamp corresponding to the center of the $k$th time bin.

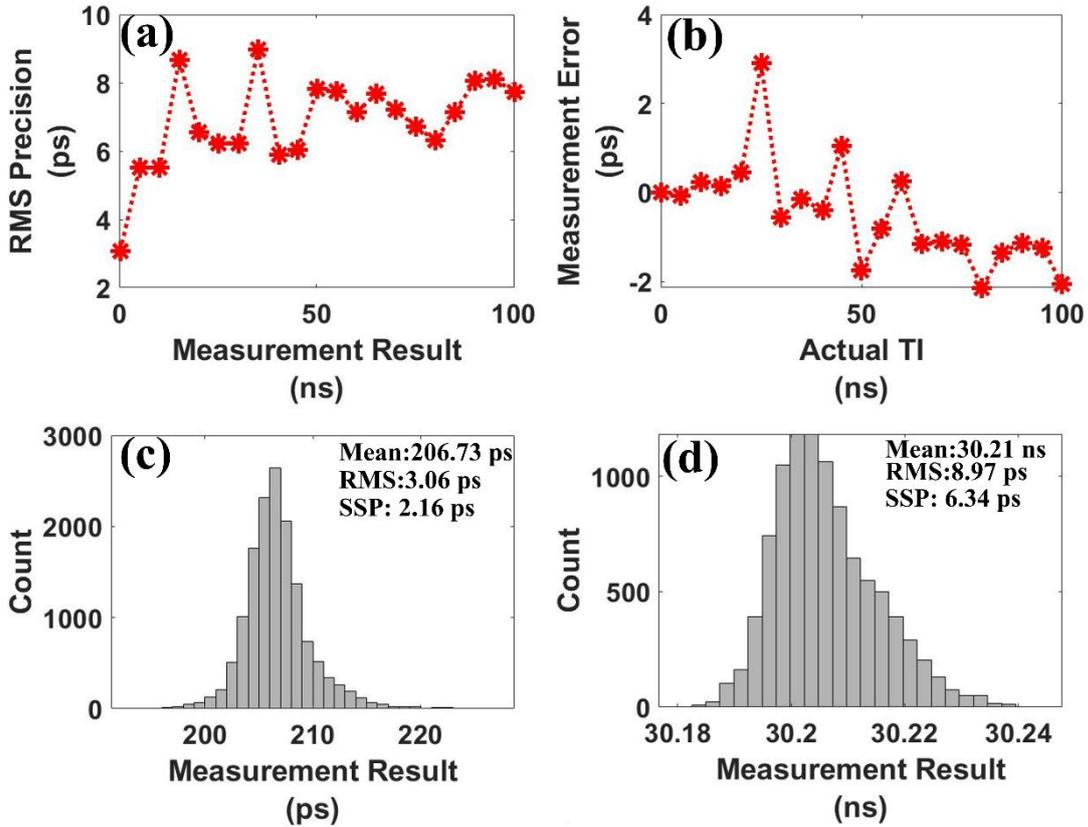

Fig. 9 (a) RMS precisions of the proposed TDC. (b) Measurement errors of the proposed TDC. (c) Measurement histogram for the TI = 0 ns. (d) Measurement histogram for the TI = 30 ns.

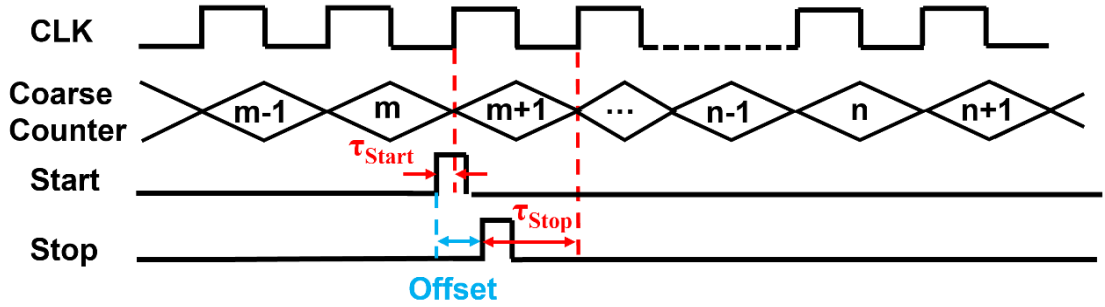

Fig. 10. The timing diagram for the measurement with the coarse counter when TI=0 ns (offset measurement).

External measurement errors and jitters are minimized for TI tests by simultaneously feeding the same signal into two channels. Then ten thousand samples are captured for each fixed TI. The RMS precision of the proposed TDC is shown in Fig. 9a. The TI varies from 0 ns to 100 ns with a 5 ns incremental step. Among all measured TIs, the best RMS precision appears when the TI equals 0 ns, achieving 3.06 ps corresponding to 2.16 ps single-shot precision (SSP, SSP=RMS/$\sqrt{2}$). Besides, the average value of measurements is 206.73 ps in this scenario, equal to the offset between two channels (Ch. start and Ch. Stop). RMS precision deteriorates with the TI increasing, achieving the worst RMS precision of 8.97 ps (corresponding to 6.34 ps SSP) at TI = 30 ns. The RMS precision fluctuates between 5 ps and 9 ps in the measured range except for TI = 0 ns. This phenomenon is caused by the jitter of the coarse counter's counting clock. In repetitive measurements for TI = 0 ns, only a few measurements are achieved with the coarse counter. And this only happens when the hit appears at the end of Ch. Start in the *m*th coarse-counting period and appears at the beginning of Ch. Stop in the *(m+1)*-th coarse-counting period due to the offset (shown in Fig. 10). However, for the rest of the TI values in TI tests, the coarse counter is always required because TIs are more than one counting period. Therefore, the RMS precision at TI = 0 ns is much better than that of the rest TIs. This phenomenon also indicates that the quality of the coarse counting clock significantly influences the measurement precision [33]. And in ultra-high-resolution TDCs (for example, a resolution of around

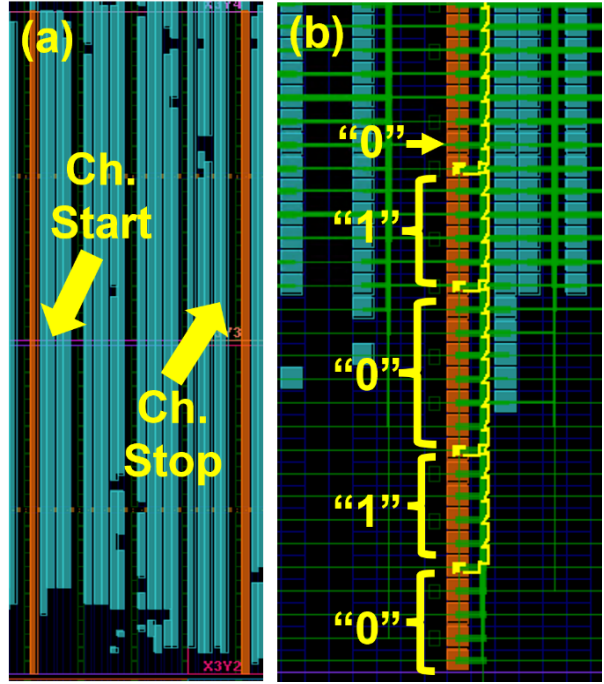

Fig.11. (a) The placement of the start and the stop channels. (b) Hardware implementation of the WU launcher.

500 fs in our design), clock jitters contribute more uncertainty to measurements than quantization errors. Besides, we analyzed measurement errors (shown in Fig. 9b) as:

$$E = (\mu - T_{offset}) - T_{actual}, \qquad (12)$$

where $T_{offset}$ is the offset between two channels, and $T_{actual}$ is the actual value of the measured TI controlled by the external signal source. Results indicate our design has less than 3 ps measurement errors in the measurement range from 5 ns to 100 ns.

## C. Hardware implementation

The hardware consumption of the proposed TDC is summarized in TABLE II. Each channel has 1920 taps from the dual-sampling TDL, consuming 234 CARRY8s, 13547 DFFs and no more than 11782 LUTs. Of them, only 120 CARRY8s, 1920 DFFs and 960 LUTs are used to construct TDLs and corresponding sampling elements in both channels. Most of the rest hardware resources are used for the bidirectional encoder. Hence, the hardware consumption of the proposed encoder is significant in the proposed TDC. However, this consumption also depends on tap numbers and implemented devices. The proposed TDC needs at least 1250 taps from the dual-sampling TDL to cover the sampling clock's whole period due to the low propagation delay of the TDL. Besides, extra 368 taps are required for wave pattern generation. They both cause the number of time bins to increase, leading to the encoder's significant hardware consumption.

Figure 11a shows the placement of the start and stop channels. We placed these two channels closely to reduce the offset. Moreover, we also routed manually for the WU launcher to ensure steady wave pattern generation. The hardware implementation of the four-edge WU launcher is shown in Fig. 11b.

## IV. Comparisons and Discussions

TABEL III summarizes recently published WU TDCs. As shown in TABLE III, most TDCs aim at the resolution from 1 ps to 10 ps, and the TDCs in Ref. [17], [34],

| | TABLE II | |
|---|---|---|
| | HARDWARE UTILIZATION FOR THE PROPOSED TDC | |
| Resource | Utilization (%) | |
| | Ch. Start | Ch. Stop |
| Tap Number | 1920 (-) | 1920 (-) |
| CARRY8 | 234 (0.81%) | 234 (0.81%) |
| DFF | 13547 (2.94 %) | 13547 (2.94 %) |
| LUT | 11773 (5.11%) | 11782 (5.11%) |

TABLE III
COMPARISON OF RECENTLY PUBLICATED WU TDCs

| Ref-year | Method | Devi. | LSB (ps) | RMS (ps) | LUT (%)[1] | DFF (%)[1] | CARRY (%)[1] | Real-time/Post Encoding |
|---|---|---|---|---|---|---|---|---|
| [15]-08 | WU-A (2 edges) | Cyclone II | 30 | 25 | - | - | - | Post Encoding |
| | WU-B | | 2.44 | 10 | - | - | - | |
| [36]-11 | WU-A (2 edges), Bin-by-bin Cali. | Virtex-4 | 6< | 8.8[2] | - | - | - | Real-time Encoding |
| [16]-11 | WU-B, Multichain-ave. | Virtex-4 | 12 | 9 | - | - | - | - |
| [17]-16 | Super WU (6 edges, 3 coding lines) | Spartan-6 | 0.90 | <6 | - | - | - | - |
| [37]-19 | WU-A (4 edges), Bin-by-bin Cali. | Kintex-7 | 2.65[3] | 3.5 | 1410[3] (1.38) | 2732[3] (1.34) | - | Real-time Encoding |
| | WU-A (8 edges), Bin-by-bin Cali. | | 1.33[3] | 3.0 | 2005[3] (1.98) | 3751[3] (1.85) | - | |
| [34]-19 | Super WU (2 edges, 8 coding lines) | Kintex-UltraScale | 0.31 | 12.32 | -[4] | -[4] | 512[3,5] (-) | Real-time Encoding |
| [35]-21 | MSWU Bin-by-bin Cali. | Kintex-7 | 0.39 | 3.30[6] | - | - | - | Post Encoding |
| [18]-22 | WU-A (2 edges), Sub-TDL, Dual-sampling | Kintex-UltraScale | 1.23 | 5.19[6] | 2460 (1.01) | 3463 (0.71) | 88 (0.29) | Real-time Encoding |
| [38]-22 | WU-A (5 edges) DSP-chain, Chunk Encoding | Artix-7 | - | 16.25 | -[7] | -[7] | -[7] | Post Encoding |
| This Work | WU-A (4 edges), Sub-TDL, Dual-sampling, Bidirectional Encoder | UltraScale+ MPSoC | 0.46 | <9 (3.06)[2] | 11773 (5.11) | 13547 (2.94) | 234 (0.81) | Real-time Encoding |

[1] Percentage of resource utilization for the implemented devices; [2] RMS precision in the best-case scenario; [3] Calculate from the literature; [4] 3200 SLICEs are used; [5] CARRY8s; [6] Value calculated from SSP; [7] 10% DSPs are used in Artix-7 XC7A200T.

[35] and the proposed TDC aim at a sub-picosecond resolution. The TDCs in Ref. [17] and [34] are based on the super WU method (combining WU A and multi-chain merging) and consume significant hardware resources. Although the multi-sampling WU method (MSWU, a technique combining WU A and WU B) in Ref. [35] can achieve a high

resolution without increasing much hardware consumption, the encoding is complex, and it is hard to perform real-time encoding in hardware platforms. Hence, our design has a trade-off between encoding complexity and hardware utilization. And it achieves real-time encoding for a high-resolution TDC with the proposed bidirectional encoder.

Compared with other four-edge WU TDCs with real-time encoding (such as that in Ref. [37]), the proposed TDC also has competitive hardware utilizing efficiency. Our design is similar to the four-edge WU TDC in Ref. [37], apart from the encoder. Due to the TDL's high propagation delay, the TDC in Ref. [37] requires 288 taps to build the WU launcher and cover the whole sampling period (with a 554 MHz sampling frequency), consuming 1410 LUTs and 2732 DFFs for each channel. By contrast, due to the TDL's low propagation delay, our design requires more than 6.5-fold taps (1920) for the WU launcher and the dual-sampling sub-TDL, consuming 8.35-fold LUTs and only 4.95-fold DFFs. The comparisons of hardware consumption between our TDC and the TDC in Ref. [37] indicate that these two designs have similar hardware utilization efficiency. However, our TDC only requires configuring LUTs for four cases (truth tables are shown in Fig. 6c), reducing complexity significantly compared with the TDC in Ref. [37].

Although our design's hardware consumption is acceptable, the hardware utilization efficiency can be further improved. For example, we can use DSP48 to sum all subsets from sub-TDLs to reduce the consumption of DFFs and LUTs. We can also multiplex adders to optimize hardware utilization. Moreover, due to the jitters of the coarse counter's clock, the proposed TDC's RMS precision deteriorates significantly when the TI is measured with the coarse counter. These two aspects still need to be improved in future work.

## V. Conclusions

Combining the dual-sampling method and the sub-TDL method, we first implemented the four-edge WU TDC in a 16 nm UltraScale+ MPSoC device. We propose a bidirectional encoder to real-time encode four-transition pseudo thermometer code and achieve a 0.46 ps resolution and less than 9 ps RMS precision with a less than 3 ps measurement error. The hardware implementation of the encoder and WU launcher is also detailed in this report. Experimental results indicate that the proposed TDC is suitable for particle physics, biomedical imaging (such as positron emission tomography, PET), and general-purpose scientific instruments.